\input harvmac
\Title{ \vbox{\baselineskip12pt\hbox{hep-th/9801139}
\hbox{HUTP-98/A002}}}
{\centerline{On N=1 Yang-Mills in Four
Dimensions }}
\centerline{Cumrun Vafa}
\centerline{ Lyman Laboratory of Physics}
\centerline{Harvard University}
\centerline{Cambridge, MA 02138, USA}
\bigskip
\noindent
Extending previous work on geometric engineering of $N=1$ Yang-Mills
in four dimensions for simply laced ($A_n,D_n,E_{6,7,8}$) gauge groups,
we construct local models for all other gauge groups
($B_n,C_n,F_4,G_2$) in terms
of F-theory.  We compute
the radius dependent superpotential upon further compactification on a circle
to $d=3$ in the dual M-theory and use it to show that the number of vacua
in four dimensions for each group is given by its dual coxeter number, in
accordance with expectations based on gaugino condensates.
\Date{January 1998}

\newsec{Introduction}

In this note we continue the previous work \ref\kva{S. Katz and C. Vafa,
Nucl. Phys. {\bf B497} (1997) 196.}\ on geometric engineering of pure $N=1$
Yang-Mills in
four dimensions and its further compactification on a circle to 3 dimensions,
by extending it to include the non-simply laced gauge groups.
In \kva\ the $N=1$ simply laced $A_n,D_n,E_{6,7,8}$
gauge groups were considered.
It was
shown there how one can compute the inequivalent vacua of these theories
in four dimensions using geometric engineering in terms of F-theory
on an elliptic Calabi-Yau fourfold
and its relation to the three dimensional description in terms of M-theory
upon further compactification on a circle.  The basic idea there was to note
that even though the four dimensional theory is strongly coupled, upon
a further compactification, the dynamics is
weakened because the theory is generically abelianized (through
vevs for Wilson lines around the circle) in which case the superpotential
of the 3 dimensional theory can be computed using point like instantons, which
are realized as
Euclidean M-theory 5-brane instantons \ref\wis{E. Witten, Nucl.Phys. {\bf B474}
(1996) 343.}.
The critical points of the superpotential for large radius
gives the number of vacua
in the four dimensional limit.  The number of vacua in four dimensions
is expected to be given by $c_2(G)$ the dual coxeter number
of the group, with the vev of gaugino bilinear $\langle \lambda^2 \rangle$
playing the role of the order parameter:
$$\langle \lambda^2 \rangle ={\rm exp}({-1\over c_2(G) g^2})\omega$$
where $\omega$ is a $c_2(G)$-th root of unity (
we have set the scale $\Lambda =1$).  This expectation
is based on one-instanton computations where there are $2c_2(G)$ gaugino
zero modes, and the cluster decomposition property of QFT's (see
\ref\revi{D. Amati, K. Konishi, Y. Meurice, G.C. Rossi, G. Veneziano,
Phys. Rep.  {\bf 162} (1988) 169.}\ for a review).

In this note we review \kva\ and indicate the
modification needed for the non-simply laced cases.  Our approach
indicates the power of geometric engineering in constructing
and studying dynamics of gauge theories.  A similar approach
for $N=4$ Yang-Mills in $d=4$ \ref\cvom{C. Vafa,
Adv. in Theor. and Math. Phys. {\bf 1} (1997) 158.}\
was used to show how Montonen-Olive duality for all gauge groups
can be reduced to T-duality of type II strings.  A similar
approach was considered for $N=2$ case in \ref\five{A. Klemm,
W. Lerche, P. Mayr, C. Vafa and N. Warner, Nucl. Phys. {\bf B477}
(1996) 746.}\ref\kkv{S.
Katz, A. Klemm and C. Vafa, Nucl. Phys. {\bf B497}
(1997) 173.}\ and led in \ref\kmv{S. Katz, P. Mayr and C. Vafa,
Adv. in Theor. and Math. Phys. {\bf 1} (1997) 53.}\
to the solution of Coulomb branch geometry for all asymptotically
free gauge theories with $SU$ gauge groups with arbitrary bifundamental
matter between pairs of groups.

\newsec{F-Theory and N=1 Yang-Mills in four dimensions}
Let us briefly review \kva :
If we consider F-theory compactification on elliptic Calabi-Yau fourfolds
we obtain an $N=1$ theory in $d=4$.
As discussed in \kva\ if we are interested in constructing an $N=1$ gauge
theory
with no matter, this can be done by considering an elliptic
fourfold which has an A-D-E singularity over a complex 2-manifold $S$
which is ``rigid'' (with $h^{1,0}=h^{2,0}=0$), such as ${\bf P}^2$
or ${\bf P}^1\times {\bf P}^1$. The rigidity is necessary
to avoid matter in the adjoint representation.
This gives rise in four dimensions to $N=1$ A-D-E Yang-Mills theory
in $4d$ where the bare gauge coupling constant is given
by
\eqn\gcc{{1\over g_4^2}=V_S}
where $V_S$ denotes the volume of $S$.
If we compactify the $N=1$ theory from $d=4$ to $d=3$ we obtain
an $N=2$ theory in $d=3$.  By the chain of duality in \ref\vf{
C. Vafa, Nucl. Phys. {\bf B469} (1996) 403.}\
the compactification of F-theory on a circle  is dual to M-theory on
the same elliptic Calabi-Yau where the radius of the circle $R$ is related
to the K\"ahler class of the elliptic fiber $V_{T^2}={1\over R}$.
Moreover there is a Weyl rescaling of the metric
so that the volume of $S$ in M-theory is given by
$$V_S^M={1\over g_3^2}={R\over g_4^2}=R\ V_S^F$$
In particular we have
\eqn\gco{{V_S^M\over R}={1\over R g_3^2}=V_S^F}
If we want
to retain the R-dependence in the physical quantities, we have
to note that the 4-fold is an elliptic one with a singularity
over the surface $S$.

$N=2$ in $d=3$ has a Coulomb branch:  The Wilson line of the
four dimensional gauge field along the circle as well
as the dual to the vector gauge field in $d=3$ which is
a scalar, form a complex scalar field $\phi$ with values
in the Cartan of the gauge group.  Going
to
non-zero value of $\phi$ (corresponding
to Wilson lines) is realized geometrically by blowing
the singularity of ADE type, and $\phi$ is identified with the blow
up parameters.  The complex part of $\phi$, being
dual to a $U(1)$ vector field, is a periodic variable.
 The
periodicity is fixed by the integrality of $H_6(K,{\bf Z})$
where $K$ is the fourfold.  In particular if the real part
of $V$ denotes
the volume of a generator of $H_6(K,{\bf Z})$, the periodicity of
its complex part
is such that the good variable is ${\rm exp}( -V)$.

For $N=2$ in $d=3$ Yang-Mills, one expects to obtain
a superpotential $W(\phi)$ as was shown for $SU(2)$
gauge group in \ref\wah{I. Affleck, J. Harvey and E.
Witten Nucl. Phys. {\bf B206} (1982)413.}. If this theory comes from a
reduction of $N=1$ in $d=4 $ on a circle
of radius $R$
where $1/g_3^2=R/ g_4^2$, the superpotential develops an $R$ dependent
piece (for the $SU(2)$ case this dependence was determined in \ref\sws{N.
Seiberg and E. Witten,
hep-th/9607163.} ).  The superpotential and its R-dependence
in the present case was determined by using the identification
of superpotential with point-like instantons corresponding to
Euclidean 5-branes \wis .  In particular it was shown in
\wis\ that for each complex 3 dimensional manifold $I$ which
is a subspace of the 4-fold and which is rigid (i.e.
where $h^{1,0}=h^{2,0}=h^{3,0}=0$) one gets a term in the
superpotential of the form ${\rm exp}(-V_I)$ where $V_I$ denotes
the superfield corresponding to the volume of $I$.

The geometry of blow up of A-D-E  is such that over each
point on $S$ we obtain a collection of $r+1$ spheres $e_i$,
where $r$ is the rank of the corresponding group.  Moreover the spheres
intersect each other according to the corresponding Affine Dynkin Diagram
(spheres correspond 1-1 to the Dynkin nodes).  Furthermore, the class of
the elliptic fiber is given by
\eqn\ellip{[T^2]=\sum_{i=1}^{r+1} a_i [e_i]}
where $a_i$ correspond to Dynkin indices of the affine
Dynkin diagram.  For $A_n$, $a_i=1$, for $D_n$, $a_i=2$
except for the four boundary nodes of the affine Dynkin
diagram where they are $1$.  For $E$-series they are given by
$$E_6:\qquad 1,1,1,2,2,2,3$$
$$E_7:\qquad 1,1,2,2,2,3,3,4$$
$$E_8:\qquad 1,2,2,3,3,4,4,5,6$$
Note that $\sum a_i=c_2(G)$ for all the A-D-E groups:
$$c_2(SU(N))=N,\quad c_2(SO(2N))=2N-2,\quad c_2(E_6)=12, \quad c_2(E_7)=18,
\quad c_2(E_8)=30$$
Recall that the volume of $T^2$ is $1/R$, thus we learn from
\ellip\ that
\eqn\rest{\sum_{i=1}^{r+1}a_i\phi_i={1\over R}}
where $\phi_i$ denotes the volume of the $i$-th sphere.  Note
that the volume of one of the nodes can be determined in terms
of all the rest.  In fact in the gauge theory description in three
dimensions all the nodes are small except for the affine node which becomes
large as $R\rightarrow 0$ and one solves \rest\ for the volume of the
affine node which becomes non-dynamical in the 3d theory.  Note that
the Dynkin number for the affine node is 1, and so we can write its volume
$V_{r+1}$ as
\eqn\gn{V_{r+1}={1\over R}-\sum_{i=1}^r a_i \phi_i }

The condition that we obtain gauge symmetry $A-D-E$ requires
the existence of a ``split'' resolution, which means that each
$e_i$ sphere which locally is a ${\bf P}^1 $ over $S$
is globally a  ${\bf P}^1$ bundle over $S$. Let us call the corresponding
3-dimensional complex manifold consisting of these ${\bf P}^1$ bundles
over $S$ by $\hat e_i$.  As was argued in \kva\ $\hat e_i$
satisfy the condition $h^{i,0}({\hat e_i})=0$ for $i\not=0$ and so give
rise to superpotential terms once they are wrapped by
Euclidean 5-branes.  We thus have $r+1$ point instantons,
one for each $\hat e_i$, i.e., one for each node of the affine Dynkin diagram.
  Thus
$$W=\sum_{i=1}^r{\rm exp}({-1\over g_3^2}\phi_i)+
{\rm exp}[{-1\over R g_3^2}+\sum_{i=1}^r {a_i\phi_i\over g_3^2}]$$
where we used the fact that $V_{\hat e_{i}}=V_{e_i}V_S^M$ and used
$V_{e_i}=\phi_i$ for $i=1,...,r$ and also used
\gco\ and \gn .
Let us define the good variables
$x_i={\rm exp}({-\phi_i\over g_3^2})$, then we can write
this as
$$W=\sum_{i=1}^r x_i+\gamma \prod_{i=1}^{r} x_i^{-a_i}$$
where $\gamma={\rm exp}({-1\over g_4^2})$.
We have rewritten the superpotential
in terms of 4d coupling (in terms of $\gamma$).
 For large
enough $R$ we should have the same number of vacua
as the 4d theory.  This in particular is the number
of critical points of $W$.  Moreover the condition that
$W$ have isolated critical points would be expected
if the 4d theory has mass gap, as is believed to be the case.  Solving
$dW=0$ we find that there are $\sum_{i=1}^{r+1}a_i=c_2(G)$
isolated critical points given by
$$x_i={\rm const.}{1\over a_i}\omega \cdot
{\rm exp}({-1\over c_2(G) g_4^2})$$
where $\omega$ is a $c_2(G)$-th root of unity.  This is as expected
based on consideration of gaugino condensates in four dimensions.
Certain aspects of these results have been further studied and
elaborated in
\ref\gom{C. Gomez and R. Hernandez, Int. Jour. Mod. Phys.
{\bf A12} (1997) 5141.}.
The case of $SU(n)$ was also studied from the field theoretic
viewpoint in \ref\int{
 O. Aharony, A. Hanany, K. Intriligator, N. Seiberg, and M.J. Strassler,
Nucl. Phys. {\bf B499} (1997) 67.}\ref\hor{J. de Boer, K. Hori and Y. Oz
Nucl.Phys. {\bf B500} (1997) 163.}.

\newsec{Non-simply Laced Case}

We now would like to extend this analysis to the non-simply
laced gauge groups.  The basic idea is that the non-simply laced
gauge groups arise from simply laced groups.  The description of
non-simply laced groups as simply laced ones modulo
the imposition of an outer automorphism is
well known mathematically, and was used in physics in
\ref\asg{P. Aspinwall and
M. Gross, Phys. Lett. {\bf B382} (1996) 81.}\ref\sixau{
M. Bershadsky, K. Intriligator,S. Kachru,
 D. Morrison, V. Sadov and C. Vafa, Nucl. Phys. {\bf B481} (1996) 215.}\cvom.
What this means in the present context
is that the blow up spheres $e_i$ are not
``split''.  In other words the $e_i$ preserve their
identity as we move over $S$ {\it only up to an outer automorphism
of the Dynkin diagram},
which exchanges some of them.  This permutation means that the
actual group is the group modulo the outer automorphism.
The non-simply laced groups are obtained as
$$SO(2n)\rightarrow SO(2n-1)\qquad {\bf Z_2}$$
$$SU(2n)\rightarrow SP(n)\qquad {\bf Z_2}$$
$$E_6\rightarrow F_4\qquad {\bf Z_2}$$
$$SO(8)\rightarrow G_2 \qquad {\bf Z_3}$$
where in the $SO$ case the ${\bf Z_2}$ outer automorphism exchanges
the two end nodes of the Dynkin diagram, in the $SU$ case the
${\bf Z_2}$ acts as a reflection on the Dynkin diagram (fixing
one node in the ordinary Dynkin diagram, and two nodes on the affine), for the
$E_6$ it exchanges the two long ends of the
Dynkin diagram and for the $SO(8)$ case the $Z_3$ cyclically
permutes the three end nodes of the ordinary Dynkin diagram
(fixing the affine node).

Let us see how this modifies the analysis of the superpotential.
For the spheres $e_i$ which are not exchanged under this
automorphism, we continue to have a well defined complex
three manifold $\hat e_i$ which is a ${\bf P}^1$ bundle
over $S$.  For the other $e_i$ which are permuted, or cyclically
exchanged, the single $\hat e_i$ does not make sense.  However
there is a double (triple) cover of the 5-brane, in the case of the ${\bf
Z_2}$ (${\bf Z_3}$) outer automorphism which does make sense and consists
of a bundle over $S$ whose fiber is the union of $e_i$ which form
a single orbit under the outer automorphism.  It is easy to see that
these fivebranes are rigid and do satisfy the criterion of
\wis\ for contributing to superpotential.  The main novelty
now is that the euclidean 5-brane volume in the case of ${\bf Z_2}$
$({\bf Z_3})$ outer automorphism is twice (three times) bigger
than what it used to be.
We thus end up with the superpotential
$$W=\sum_{i=1}^{r'}x_i^{m_i} +\gamma \prod_{i=1}^{r'} x_i^{-m_ia_i}$$
where $r'$ denotes the rank of the non-simply laced group and $m_i$ denotes
the number of nodes in the same orbit as $e_i$ under the outer
automorphism.   However, we have to note that the good variables
now are
$$y_i=x_i^m.$$
The reason for this, as explained before is that $H_6(K,{\bf Z})$
fixes the periodicity of the phase of the chiral
fields, where $K$ is the Calabi-Yau fourfold.
In the non-simply laced case some of the generators of $H_6(K,{\bf Z})$
are $m_i$ times bigger than what they used to be.  This implies
that $y_i$ are now the correct variables, in terms of which we have
$$W=\sum_{i=1}^{r'}y_i +\gamma \prod_{i=1}^{r'}y_i^{-a_i}$$
This is now exactly of the same form as in the simply laced
case and so finding the critical points
of the superpotential in this case gives
the number of vacua which is
$$1+\sum_{i=1}^{r'}a_i=c_2(G')$$
where the sum is now over all the Dynkin indices of the
simply laced group, one for each orbit of the outer automorphism.
It is easy to check that this gives for the number of vacua
the dual coxeter number of the group, i.e.,
$$c_2(SO(2n-1))=2n-3,\quad c_2(SP(n))=n+1,\quad c_2(G_2)=4,
\quad c_2(F_4)=9$$
(for the case of $F_4$ the inequivalent $a_i$
are $1,2,2,3$).

\vglue 2cm

We would like to thank  A. Johansen and S. Katz
for valuable discussions.

This research was supported in part by
NSF grant PHY-92-18167.

\listrefs
\end